\documentclass[aps,letter,prb,twocolumn]{revtex4}
\usepackage{graphics,graphicx}
\usepackage{color}
\usepackage{wrapfig}
\usepackage{enumerate}
\usepackage{amssymb,amsmath}
\usepackage{bm}
\usepackage[normalem]{ulem} % added by Keola to support \sout{text goes here} command

\begin{document}
\title{Strange correlations in spin-1 Heisenberg antiferromagnets}
\author{Keola~Wierschem$^{1}$ and Pinaki~Sengupta$^{1}$}
\affiliation{$^{1}$School of Physical and Mathematical Sciences, Nanyang Technological University, 21 Nanyang Link, Singapore 637371}
\date{\today}
\begin{abstract}
We study the behavior of the recently proposed ``strange correlator'' [Phys. Rev. Lett. {\bf 112}, 247202 (2014)] in spin-1 Heisenberg antiferromagnetic chains with uniaxial single-ion anisotropy. Using projective quantum Monte Carlo, we are able to directly access the strange correlator in a variety of phases, as well as to examine its critical behavior at the quantum ph=ase transition between trivial and non-trivial symmetry protected topological phases. After finding the expected long-range behavior in these two symmetry conserving phases, we go on to verify the topological nature of two-leg and three-leg spin-1 Heisenberg antiferromagnetic ladders. This demonstrates the power of the strange correlator in distinguishing between trivial and non-trivial symmetry protected topological phases.
\end{abstract}
\maketitle

\section{Introduction}

The study of many body effects like the Haldane phase and the fractional quantum Hall effect led to the discovery of an entire class of quantum mechanical ground states -- topologically ordered phases -- that fall outside the standard Landau framework. Unlike conventional states of matter, topological phases are not broken-symmetry ground states characterizable by a local order parameter. Instead, they have an underlying topological structure that distinguishes them from disordered (topologically trivial) phases. A finite gap separates the lowest excitations from the ground state in the bulk, but there exist one or more gapless edge states, which is the defining characteristic of this class of phases. The theoretical prediction and subsequent experimental discovery of topological (band) insulators have ushered in a period of heightened interest in topological phases.

While many experimentally realized topological phases, such as fractional quantum Hall states and topological insulators, can be described within the framework of non-interacting electrons, a natural question that arises in the study of these phases is the role of interactions. To address this, a minimal generalization of the free fermion topological phase, known as the symmetry protected topological (SPT) phase, was proposed. An SPT state is defined as the ground state of an interacting many body system that is comprised of a gapped bulk state that preserves all the symmetries of the system and a gapless non-trivial edge state that is protected by one or more symmetry. In keeping with its minimal character, phases with long-range topological order (defined by long-range entanglement) are excluded from the SPT classification. Instead, SPT states are characterized by short-range entanglement. Significant progress has been made over the past few years in the understanding of SPT states. This includes a formal mathematical classification of these states,~\cite{Gu2009,Chen2013} as well as detailed investigations of proposed SPT phases (for both interacting fermions and bosons). The relative simplicity allows us to understand the emergence of topological phases from the interplay of strong correlations, symmetry and topology in these systems and, in turn, provides deeper insight into more complex topological states such as spin liquids and non-Fermi liquid metals.

While several SPT states have been discovered and studied in detail, only a handful of microscopic Hamiltonians with SPT ground states are known to date. Even more worryingly, given a Hamiltonian, there exists no well-established probe to determine if the ground state has SPT character. Although a degenerate entanglement spectrum is often used as an indicator of SPT order,~\cite{Pollmann2010} this method may fail to correctly identify SPT phases protected by an off-lattice symmetry. Additionally, the relation between the low-lying entanglement spectrum and the ground state wave function has been called into question.~\cite{Chandran2013} Recently, a strange correlator has been proposed as a direct probe of the SPT character of a wave function and has been demonstrated to identify some well known SPT phases successfully.~\cite{You2014} In this work, we shall present details on how to evaluate the strange correlator in quantum Monte Carlo (QMC) simulations and use it to probe the topological nature of the ground state of spin-1 Heisenberg chains and ladders.

%This paper is arranged as follows. In section~\ref{model} we describe the spin model, while in section~\ref{methods} we describe our projective QMC implementation and how it can be used to calculate the strange correlator of You {\it et al.}~\cite{You2014} We describe our results in section~\ref{results}, which is followed by a discussion in section~\ref{discussion}. Finally, we conclude our paper in section~\ref{conclusion}.

\section{Model}\label{model}

The Haldane phase of the spin-1 Heisenberg antiferromagnetic chain remains the earliest and most well-understood of all interacting SPT phases. This simple model has a surprisingly rich ground state phase diagram. In addition to the Haldane phase, a topologically trivial quantum paramagnetic phase (the so-called large-$D$ phase) appears when strong uniaxial easy-plane single-ion anisotropy is introduced. For spin-1 Heisenberg antiferromagnetic chains couped into a ladder geometry, the topological nature of the ground state exhibits an even/odd effect: trivial SPT character for even-leg ladders and non-trivial SPT character for odd-leg ladders. Our goal is to demonstrate that the strange correlator correctly identifies the varying SPT character of these respective ground states, as well as to examine the critical behavior of the strange correlator at the quantum phase transition between the Haldane and large-$D$ phases.

\begin{figure}
\centering
\includegraphics[clip,trim=0cm 0cm 0cm 0cm,width=\linewidth]{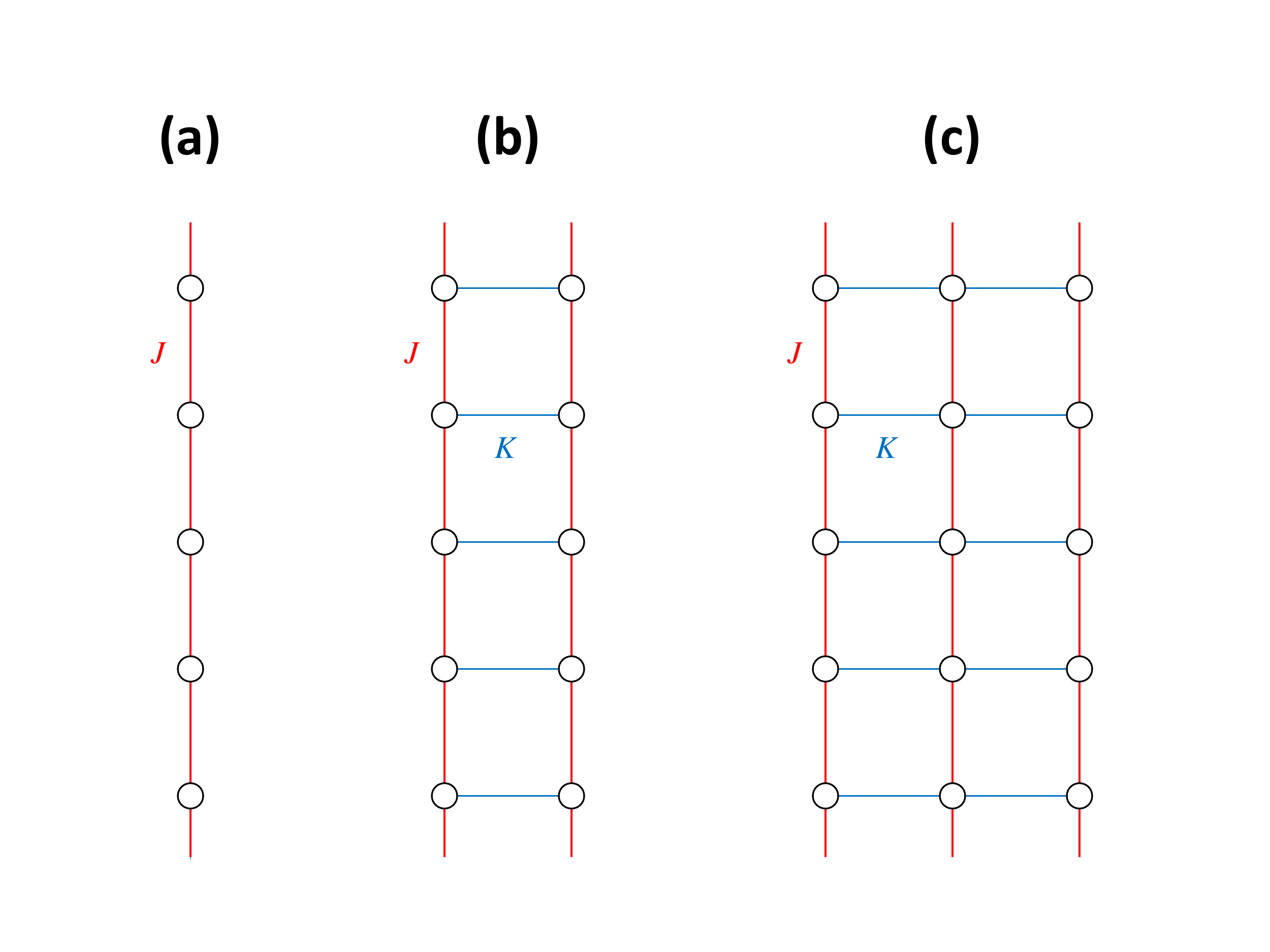}
\caption{(Color online) Illustration of the three lattice geometries considered in this work: {\bf (a)} chain, {\bf (b)} two-leg ladder, and {\bf (c)} three-leg ladder. The spin exchange couplings $J$ and $K$ are shown as red and black lines, respectively. Empty black circles represent lattice sites.}
\label{geometry}
\end{figure}

To this end, we study spin-1 Heisenberg antiferromagnetic chains and ladders with uniaxial single-ion anisotropy,  described by the Hamiltonian
\begin{equation}
\label{hamiltonian}
{\cal H}=J\sum_{\left<ij\right>}\vec{S_{i}}\cdot\vec{S_{j}}+K\sum_{\left[ij\right]}\vec{S_{i}}\cdot\vec{S_{j}}+D\sum_{i}\left(S_{i}^{z}\right)^2.
\end{equation}
Here, $\left<ij\right>$ refers to neighbors along a chain, while $\left[ij\right]$ refers to neighbors between adjacent chains (see Fig.~\ref{geometry}). In the following, we set the spin exchange coupling $J$ to unity, thereby defining the energy scale of our system. This leaves the interchain coupling $K$ and single-ion anisotropy $D$ as our only Hamiltonian parameters. For the ladder geometries, we further set $D=0$, while for the chain geometry the parameter $K$ becomes meaningless. We use QMC methods to study finite-size systems of size $N$ and length $L$, where $N=L$, $2L$ and $3L$ for chain, two-leg ladder and three-leg ladder geometries, respectively. Periodic boundary conditions are employed along the length of the system.

\section{Methods}\label{methods}

To investigate the above model in the ground state limit, we use a projective variant of the stochastic series expansion QMC method.\cite{Sandvik1991,Sandvik1992} The main idea is as follows: instead of expanding the density matrix as a Taylor series of Hamiltonian operations, we project out the ground state by repeated Hamiltonian operation on a trial wave function. While the presence of a trial wave function explicitly removes the usual periodicity in the imaginary-timelike dimension of the operator string, it can be thought of as a set of vertices of infinite weight. Thus, we can still utilize the directed loop equations of Sylju\r{a}sen and Sandvik,\cite{Syljuasen2002} minimize bounce probabilities in the loop algorithm, and obtain efficient global updates.

Now we describe our projective QMC scheme in detail. First, let us examine the effect of $m$ repeated Hamiltonian operations on a trial wave function,
\begin{equation}
\left({\cal H}- {\cal C}\right)^{m}|\psi\rangle=\left({\cal H}-{\cal C}\right)^{m}\sum_{\alpha}c_{\alpha}|\alpha\rangle.
\end{equation}
Here, we have expanded $|\psi\rangle$ in the basis of energy eigenstates $|\alpha\rangle$ with coefficients given by $c_{\alpha}=\langle\alpha|\psi\rangle$. The constant ${\cal C}$ is chosen to make $\left({\cal H}-{\cal C}\right)$ negative definite. Thus, as long as $c_{0}\neq0$, the projection of $|\psi\rangle$ will be dominated by the ground state terms,
\begin{equation}
\left({\cal H}-{\cal C}\right)^{m}|\psi\rangle=\sum_{\alpha}c_{\alpha}\left(E_{\alpha}-{\cal C}\right)^{m}|\alpha\rangle.
\end{equation}
This can be made explicit by rewriting the expression as
\begin{equation}
\left(\frac{{\cal H}-{\cal C}}{E_{0}-{\cal C}}\right)^{m}|\psi\rangle=\sum_{\alpha}c_{\alpha}\left(\frac{E_{\alpha}-{\cal C}}{E_{0}-{\cal C}}\right)^{m}|\alpha\rangle.
\end{equation}
The ground state is approached as $m\rightarrow\infty$. In addition to the power-based projector $\left({\cal H}-{\cal C}\right)^m$, it is also possible to use an exponential projector $\exp\left[-\beta\left({\cal H}-{\cal C}\right)\right]$. A detailed description of the exponential projector has been given by Farhi {\it et al.}\cite{Farhi2012}

Having a valid ground state projector, we can evaluate ground state observables as
\begin{equation}
\langle{\cal O}\rangle=\frac{\langle\psi|\left({\cal H}-{\cal C}\right)^{m}{\cal O}\left({\cal H}-{\cal C}\right)^{m}|\psi\rangle}{\langle\psi|\left({\cal H}-{\cal C}\right)^{2m}|\psi\rangle}.
\end{equation}
These observables are calculated at the ``middle'' of the (open) operator string. Appropriate sampling weights can be derived by expanding the projector $\left({\cal H}-{\cal C}\right)^m$ as a summation over all possible operator strings. Each operator string consists of a product of $2m$ bond operators. This is completely analogous to the corresponding expansion of the density matrix into a summation over operator strings in the standard stochastic series expansion QMC technique,~\cite{Sandvik1991,Sandvik1992} with the added simplicity that the length of our operator string remains fixed due to our choice of a power-based projector.

Within the same formulation, we can also easily compute overlap of the ground state wave function with an arbitrary wave function $|\Omega\rangle$, as required for calculating the so-called strange correlator~\cite{You2014}
\begin{equation}
C(r-r')=\frac{\langle\Omega|\phi(r)\phi(r')\left({\cal H}-{\cal C}\right)^{2m}|\psi\rangle}{\langle\Omega|\left({\cal H}-{\cal C}\right)^{2m}|\psi\rangle}.
\end{equation}
Here, $\phi(r)$ is a local operator that for the present work we set to $\phi(r)\phi(r')=S_{r}^{+}S_{r'}^{-}+S_{r}^{-}S_{r'}^{+}$. When $|\Omega\rangle$ is a trivial symmetric product state, You {\it et al.}~\cite{You2014} have shown that the strange correlator must be long ranged if the ground state $|\Psi\rangle=\left({\cal H}-{\cal C}\right)^{2m}|\psi\rangle$ is a non-trivial SPT state. Interestingly, within our projective QMC formulation, the strange correlator is simply a normal correlation function measured at the ``ends'' of the operator string. This correspondence highlights the physical interpretation of the strange correlator as a correlation function at the temporal boundary of the time-evolved ground states $|\Psi\rangle$ and $|\Omega\rangle$.\cite{You2014}

In the current work, we choose both the trivial product state and the trial wave function to be the product state of zero spin projection along the $z$-axis,
\begin{equation}
|\psi\rangle=|\Omega\rangle=\prod_i|0\rangle_i.
\end{equation}
This choice conserves the on-site symmetry of the Hamiltonian in Eq.~\ref{hamiltonian}. Also, note that in combination with the choice of the local operator $\phi(r)$, this definition has the convenient feature that $C(0)=2$. Another convenience arises when taking into account the usual sublattice rotation of $\pi$ along the $z$-axis of the spin space that is needed to ensure negative-definite off-diagonal vertex weights in the bond expansion of the Hamiltonian. For a projective QMC implementation, this sublattice rotation should also be applied to the trial wave function. Here we note that the current choice of $|\psi\rangle$ is invariant under such a transformation. This transformation introduces a shift of $\pi$ in the momentum of spin excitations. Henceforth, we assume that the momentum vector is measured from the shifted point, i.e. $k\rightarrow k+\pi$.

In order to improve the statistical sampling of the strange correlator at the ends of the operator string, as well as normal observables at the middle level of the operator string, we introduce a bias in our loop updates to preferentially start the loop at these levels of the operator string. Satisfaction of detailed balance is contingent upon equal probabilities of starting forward and reverse loops, which is not effected by our added bias. This is because loops starting at different levels of the operator string are not connected to each other.\cite{Syljuasen2002}

In the next section, we proceed to analyze the strange correlator in the ground state thermodynamic limit. Using system sizes of length $16\le L\le96$, we are able to accurately determine the proper scaling limits of the strange correlator in several symmetric phases. Analogous to the ground state scaling of the operator string length in finite-temperature stochastic series expansion QMC, we choose $m\propto L^{2}$ to converge to the ground state.

\section{Results}\label{results}

This section is organized as follows. First, we consider the behavior of the strange correlator in the spin-1 Heisenberg antiferromagnetic chain with uniaxial single-ion anisotropy. This system has both trivial and non-trivial SPT states -- the large-$D$ and Haldane phases, respectively. These two phases are separated by a continuous phase transition, which allows for the investigation of critical behavior of the strange correlator. Subsequently, we examine the behavior of the strange correlator in the spin-1 Heisenberg antiferromagnetic two-leg and three-leg ladders. Finally, we look at the finite-size scaling behavior of an order parameter associated with the strange correlator.

\subsection{Chain}

Following Haldane's seminal conjecture that the ground state of integer spin Heisenberg antiferromagnets in one dimension should be gapped,~\cite{Haldane1983a,Haldane1983b} it was quickly realized that the ground state of the spin-1 Heisenberg antiferromagnetic chain is closely related to the exact ground state (the so-called AKLT state) of a Hamiltonian constructed by a sum of spin-2 projectors on each bond.~\cite{Affleck1987} The AKLT state has the interesting property of possessing fractionalized spin-1/2 degrees of freedom at both ends of an open chain. This topological character can be captured by the string order parameter,~\cite{denNijs1989} which is also present in the Haldane phase of the spin-1 Heisenberg antiferromagnetic chain. In other words, the Haldane phase is adiabatically connected to the AKLT state. In the presence of uniaxial single-ion anisotropy, there is a distinct topologically trivial quantum paramagnetic phase for strong easy-plane values of the anisotropy -- the so-called large-$D$ phase. The Haldane and large-$D$ phases are separated by a Gaussian phase transition at the critical value $D_c\approx0.97$.

\begin{figure}
\centering
\includegraphics[clip,trim=0cm 0cm 0cm 0cm,width=\linewidth]{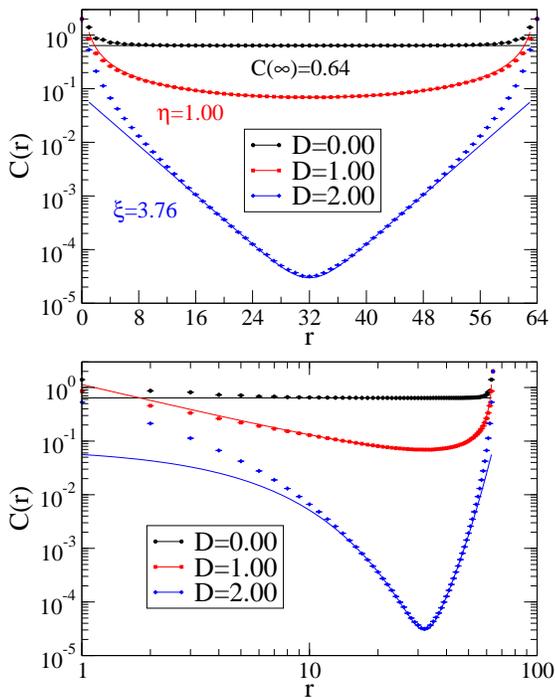}
\caption{(Color online) Strange correlator in the Haldane ($D=0$) and large-$D$ ($D=2$) phases, as well as in the vicinity of the critical point ($D=1$). The long-range order and exponential decay in the Haldane and large-$D$ phases, respectively, are readily apparent in the log-normal plot of the upper panel, while the lower panel illustrates algebraic decay near $D_c$ using a log-log plot. Solid lines are fits to $C(r)=C(\infty)$, $C(r)\sim r^{-\eta}+(L-r)^{-\eta}$, and $C(r)\sim e^{-r/\xi}+e^{-(L-r)/\xi}$ in the region $L<4r<3L$ (black, red and blue lines, respectively). Taking $J=1$, we find $2m=L^2$ is sufficient to converge to the ground state limit.}
\label{correlation}
\end{figure}

In Fig.~\ref{correlation}, we plot the strange correlator as a function of distance for a 64-site chain. Results are shown for values of the single-ion anisotropy in the Haldane and large-$D$ phases, as well as in the vicinity of the critical point $D_c=0.97$ separating these two phases. By fitting the strange correlator to appropriate forms in the region $L<4r<3L$, we confirm that it indeed approaches a constant value $C(\infty)=0.64$ in the non-trivial Haldane phase, while in the trivial large-$D$ phase it scales exponentially to zero with correlation length $\xi=3.76$. Near the critical point, the correlation decays algebraically with a power $\eta=1.00$.

\subsection{Two-leg ladder}

The two-leg spin-1 Heisenberg ladder (with $D=0$) has been shown to have a topologically trivial SPT ground state.~\cite{Pollmann2012} This can be understood by considering the limiting case of strong rung interactions, $J\ll K$. In this limit, the ground state is well approximated as a product of rung singlets. By definition, such a state is topologically trivial. When coupled with results from a previous QMC study that demonstrated a lack of any phase transition,~\cite{Todo2001} this leads to the conclusion that the ground state of the isotropic two-leg spin-1 Heisenberg ladder is always topologically trivial.

\begin{figure}
\centering
\includegraphics[clip,trim=0cm 0cm 0cm 0cm,width=\linewidth]{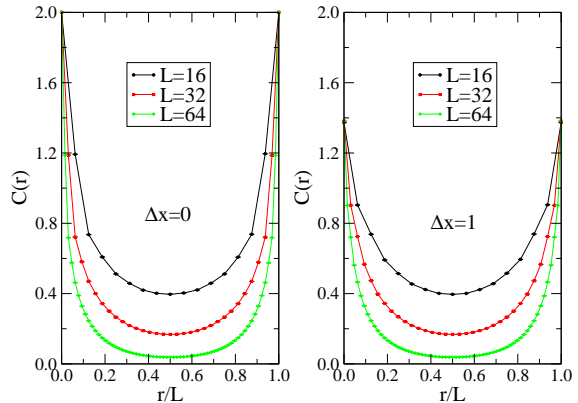}
\caption{(Color online) Strange correlator in the two-leg spin-1 ladder with $D=0$ and $J=K=1$. We find $2m=L^2$ is sufficient to converge to the ground state limit.}
\label{ladder}
\end{figure}

Here, we calculate the strange correlator for antiferromagnet rung interactions with $J=K=1$. As seen in Fig.~\ref{ladder}, the strange correlator at long distances quickly decays to zero as our system size approaches the thermodynamic limit. This is true both for correlations within a single leg of the ladder ($\Delta x=0$) as well as for correlations between opposite legs of the ladder ($\Delta x=1$). Thus, we see that the strange correlator correctly identifies the ground state of the spin-1 Heisenberg antiferromagnetic two-leg ladder as a topologically trivial state.

\subsection{Three-leg ladder}

\begin{figure}
\centering
\includegraphics[clip,trim=0cm 0cm 0cm 0cm,width=\linewidth]{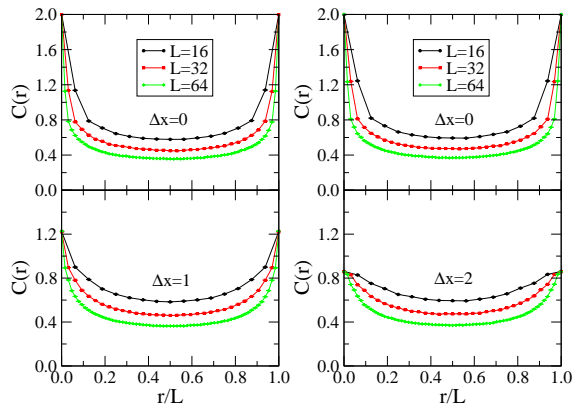}
\caption{(Color online) Strange correlator in the three-leg spin-1 ladder with $D=0$ and $J=K=1$. Correlations involving the middle leg of the ladder are shown in the left panels, while those involving only the outer legs are shown in the right panels. We find $2m=3L^2$ is sufficient to converge to the ground state limit.}
\label{tube}
\end{figure}

In contrast to the two-leg spin-1 ladder, the ground state of the three-leg spin-1 ladder is a topologically non-trivial SPT state.\cite{Charrier2010} This can be seen in Fig.~\ref{tube}, where strange correlator converges to a non-zero value as our system size approaches the thermodynamic limit. As before, the long distance behavior is the same for correlations within a single leg of the ladder ($\Delta x=0$, inner or outer legs) as well as for correlations between different legs of the ladder ($\Delta x=1$ or $\Delta x=2$). Once again, we see that the strange correlator correctly identifies the ground state SPT character, this time finding a non-trivial SPT ground state for the spin-1 Heisenberg antiferromagnetic three-leg ladder.

\subsection{Finite-size scaling}

In order to make a more quantitative statement on the scaling properties of the strange correlator in the thermodynamic limit, we define a finite-size scaling parameter $\Psi_L=\frac{1}{N}\sum_{\vec{r}}C(\vec{r})$ based on the corresponding static structure factor (see Eq.~\ref{parameter} in the appendix). In Fig.~\ref{order}, we show the system size dependence of $\Psi_L$ for various SPT phases. As expected, within the non-trivial SPT phases (chain and three-leg ladder with $D=0$), we find $\Psi_L$ approaches a constant value exponentially with system size. This is confirmed by fitting $\Psi_L$ using the scaling form of Eq.~\ref{exponential} that we derive in the appendix. For the trivial SPT phases (chain with $D=2$, two-leg ladder with $D=0$), $\Psi_L$ instead scales exponentially to zero with system size. Excellent fits are obtained using Eq.~\ref{exponential} with $\Psi_{\infty}=0$. In the last scenario, we fit $\Psi_L$ near the critical point (chain with $D=1$) using the scaling form of Eq.~\ref{algebraic}. This confirms that the strange correlator follows the usual critical behavior at the phase boundary between trivial and non-trivial SPT phases.

\begin{figure}
\centering
\includegraphics[clip,trim=0cm 0cm 0cm 0cm,width=\linewidth]{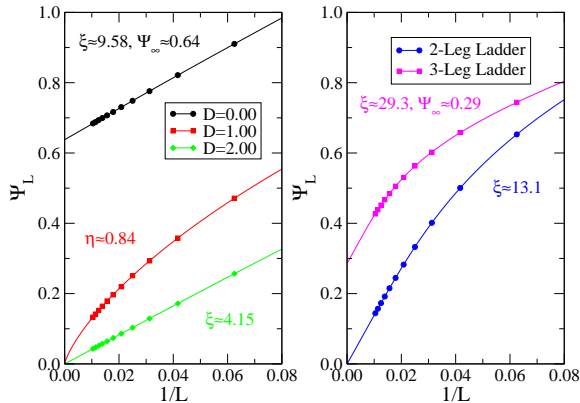}
\caption{(Color online) Finite-size scaling of the order parameter for the strange correlator in chain (left panel) and ladder (right panel) geometries. Lines are obtained as fits to the data using the finite-size scaling forms derived in the appendix. Parameters are the same as in earlier figures.}
\label{order}
\end{figure}

The critical behavior at the quantum phase transition between the Haldane and large-$D$ phases is captured by a conformal field theory that maps onto a free Gaussian model.~\cite{DegliEspostiBoschi2003}
The critical exponents for this Gaussian transition can be expressed in terms of a single parameter $K$ (the Luttinger parameter). For the Green's function $G(r)=\langle S_0^xS_r^x+S_0^yS_r^y\rangle$ the anomalous dimensionality should be given by $\eta=1/2K$, while the critical exponent governing the correlation length is expected to be $\nu=1/(2-K)$. Similar relations can be derived through the well-known bosonization technique.

\begin{figure}
\centering
\includegraphics[clip,trim=0cm 0cm 0cm 0cm,width=\linewidth]{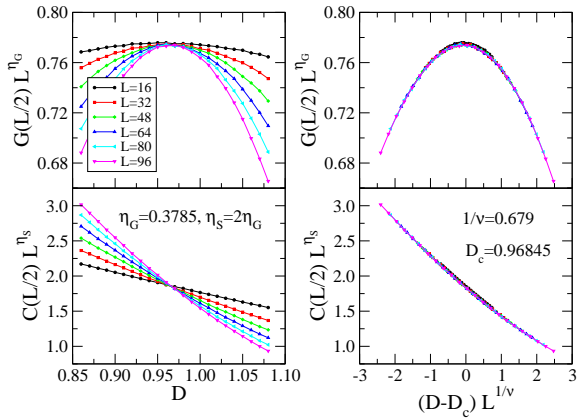}
\caption{(Color online) Finite-size scaling of the order parameter across the Gaussian critical point of the Heisenberg AFM chain with single-ion anisotropy for the normal Green's function (top panels) and the strange correlator (bottom panels). Lines are shown as guides to the eye. We find $2m=4L^2$ is sufficient to converge to the ground state limit.}
\label{fss}
\end{figure}

The Gaussian transition has been well-studied,~\cite{DegliEspostiBoschi2003,Tzeng2008a,Tzeng2008b,Albuquerque2009,Hu2011} with recent results from the density matrix renormalization group obtaining $D_c=0.96845(8)$ and $K=1.321(1)$.~\cite{Hu2011} To test our method, in the top panels of Fig.~\ref{fss} we plot the Green's function at half the system size, $G(L/2)$, multiplied by $L^\eta$ to obtain a dimensionless parameter. As can be seen in the left panel, this dimensionless parameter is independent of the system size at the critical point $D_c$, while the right panel demonstrates finite-size data collapse. Interestingly, when we plot the strange correlator using an exponent $\eta=2\eta_G$ (lower panels, Fig.~\ref{fss}), we see a similar curve crossing and finite-size collapse! This gives a value $\eta=0.757$ that is lower than our estimate from fitting $\Psi_L$ in Fig.~\ref{order}, yet reasonably close considering the distance of $D=1$ from $D_c$. It remains to be seen why this relation works so well here, and whether or not it applies generally at the boundary between two distinct SPT phases.

\section{Discussion}\label{discussion}

In one dimension, a non-local order parameter can be defined to distinguish between trivial and non-trivial SPT phases.~\cite{Pollmann2012b} For the Haldane phase of the spin-1 Heisenberg antiferromagnetic chain, this is non other than the original string order parameter.~\cite{denNijs1989} Similar non-local order parameters can also be defined on two-leg~\cite{Todo2001} and three-leg~\cite{Charrier2010} ladders. However, note that such non-local order parameters can either measure hidden symmetry conservation or hidden symmetry breaking, depending upon their construction.~\cite{Pollmann2012b} Thus, the four-body string order parameter of Todo {\it et al.}~\cite{Todo2001} is in fact evidence for a trivial SPT state in the two-leg ladder.

While carefully constructed string order parameters may allow for the characterization of SPT phases in one dimension, they are only valid in the presence of dihedral symmetry (${\cal D}_{2}$ or ${\cal Z}_{2}\times{\cal Z}_{2}$).~\cite{Pollmann2012b} Additionally, they do not extrapolate easily to higher dimensions. The strange correlator is thus a powerful tool for the generic investigation of SPT properties regardless of on-site symmetry, and extends easily to higher dimensions. As demonstrated by You {\it et al.},~\cite{You2014} the strange correlator can easily distinguish trivial and non-trivial SPT states in one and two dimensions. In three dimensions, a long-range or quasi-long-range strange correlator still implies non-trivial SPT order; however, due to the possiblity of topologically ordered edge states, it becomes possible for a non-trivial SPT state to have short-range strange correlator.~\cite{You2014}

\section{Conclusion}\label{conclusion}

In conclusion, we have implemented a projective QMC method that is able to calculate the strange correlator in a wide variety of phases. Using this method to study the spin-1 Heisenberg antiferromagnetic chain, we have verified the topological nature of this prototypical SPT system. Adding a uniaxial single-ion anisotropy, we find evidence of critical behavior in the strange correlator at the quantum phase transition between the Haldane and large-$D$ phases. Thus, the strange correlator can be used as an order parameter for phase transitions between trivial and non-trivial SPT states. We have also calculated the strange correlations in two-leg and three-leg ladders to verify their relative trivial and non-trivial SPT phases. Although the topological characterization of these phases was known from past work, the QMC methods implemented here are easily extended to higher dimensions. Further, the strange correlator should continue to maintain at least quasi-long-range behavior in two dimensions.~\cite{You2014} This paves the way for applications to systems in two dimensions, where string order becomes ill defined.\cite{Wierschem2014}

\begin{acknowledgments}
We thank Anders Sandvik and Yi-Zhuang You for helpful discussions. We also thank Cenke Xu for useful comments. This research used resources of the National Energy Research Scientific Computing Center, which is supported by the Office of Science of the U.S. Department of Energy under Contract No. DE-AC02-05CH11231.
\end{acknowledgments}

\appendix*

\section{Finite-Size Scaling Forms}

Here we derive finite-size scaling forms for the order parameter, which can be defined as the structure factor per site
\begin{equation}
\label{parameter}
\Psi_L=\frac{S(0)}{N}=\frac{1}{N}\sum_{\vec{r}}C(\vec{r}).
\end{equation}
This expression can be converted into an integral in the continuum limit that is obtained as $L\rightarrow\infty$:
\begin{equation}
S(0)=\int_0^LC(r)dr.
\end{equation}
Assuming that the correlation function approaches the asymptotic form $C(r)\sim e^{-r/\xi}$ for $r>R$, we replace the integral from $0$ to $R$ by a ``core charge'' $C_0$. This yields a simple expression for the structure factor,
\begin{equation}
S(0)=C_0+A\int_R^{L-R}\left[e^{-r/\xi}+e^{-(L-r)/\xi}\right]dr,
\end{equation}
where we take into account the periodicity of $C(r)$ for finite simulation cells with periodic boundary conditions. This gives
\begin{equation}
S(0)=C_0+A\xi\left[e^{-R/\xi}-e^{-(L-R)/\xi}\right],
\end{equation}
which ultimately is the same as
\begin{equation}
S(0)=a-be^{-L/\xi},
\end{equation}
i.e. we cannot uniquely determine $C_0$ and $R$. This leads to finite-size scaling form
\begin{equation}
\label{exponential}
\Psi_L=\Psi_{\infty}+\left(a-be^{-L/\xi}\right)/L,
\end{equation}
where $\Psi_{\infty}\neq0$ allows for an exponential decay to a non-zero value, as is the case for ordered phases.

At a critical point, the expected asymptotic scaling form becomes $C(r)\sim r^{-\eta}$, so instead we find
\begin{equation}
S(0)=C_0+A\int_R^{L-R}\left[r^{-\eta}+(L-r)^{-\eta}\right]dr,
\end{equation}
which after integration becomes
\begin{equation}
S(0)=C_0+\frac{A}{1-\eta}\left[\left(L-R\right)^{1-\eta}-R^{1-\eta}\right].
\end{equation}
Again, we cannot uniquely determine $C_0$ and $R$, which leaves the general scaling form
\begin{equation}
S(0)=a+b\left(L-R\right)^{1-\eta}.
\end{equation}
For small $R/L$ this can be replaced by the simpler form
\begin{equation}
S(0)=a+bL^{1-\eta}.
\end{equation}
Thus, we have used the following finite size scaling form for the order parameter near the critical point:
\begin{equation}
\label{algebraic}
\Psi_L=a/L+b/L^{\eta}.
\end{equation}

The above finite-size scaling forms can also be derived for higher dimensions $d>1$ by using the relation $\Psi_{L}=S(0)L^{-d}$. In this case, the correlation function at a critical point is defined as $C(r)\sim r^{2-(d+z+\eta)}$, where $z$ is the dynamic critical exponent for ground state phase transitions. For the Gaussian phase transition separating the Haldane and large-$D$ phases, we expect $z=1$.~\cite{Albuquerque2009}

\bibliographystyle{apsrev}
\bibliography{strange-ref}

\begin{thebibliography}{23}
\expandafter\ifx\csname natexlab\endcsname\relax\def\natexlab#1{#1}\fi
\expandafter\ifx\csname bibnamefont\endcsname\relax
  \def\bibnamefont#1{#1}\fi
\expandafter\ifx\csname bibfnamefont\endcsname\relax
  \def\bibfnamefont#1{#1}\fi
\expandafter\ifx\csname citenamefont\endcsname\relax
  \def\citenamefont#1{#1}\fi
\expandafter\ifx\csname url\endcsname\relax
  \def\url#1{\texttt{#1}}\fi
\expandafter\ifx\csname urlprefix\endcsname\relax\def\urlprefix{URL }\fi
\providecommand{\bibinfo}[2]{#2}
\providecommand{\eprint}[2][]{\url{#2}}

\bibitem[{\citenamefont{Gu and Wen}(2009)}]{Gu2009}
\bibinfo{author}{\bibfnamefont{Z.-C.} \bibnamefont{Gu}} \bibnamefont{and}
  \bibinfo{author}{\bibfnamefont{X.-G.} \bibnamefont{Wen}},
  \bibinfo{journal}{Phys. Rev. B} \textbf{\bibinfo{volume}{80}},
  \bibinfo{pages}{155131} (\bibinfo{year}{2009}).

\bibitem[{\citenamefont{Chen et~al.}(2013)\citenamefont{Chen, Gu, Liu, and
  Wen}}]{Chen2013}
\bibinfo{author}{\bibfnamefont{X.}~\bibnamefont{Chen}},
  \bibinfo{author}{\bibfnamefont{Z.-C.} \bibnamefont{Gu}},
  \bibinfo{author}{\bibfnamefont{Z.-X.} \bibnamefont{Liu}}, \bibnamefont{and}
  \bibinfo{author}{\bibfnamefont{X.-G.} \bibnamefont{Wen}},
  \bibinfo{journal}{Phys. Rev. B} \textbf{\bibinfo{volume}{87}},
  \bibinfo{pages}{155114} (\bibinfo{year}{2013}).

\bibitem[{\citenamefont{Pollmann et~al.}(2010)\citenamefont{Pollmann, Turner,
  Berg, and Oshikawa}}]{Pollmann2010}
\bibinfo{author}{\bibfnamefont{F.}~\bibnamefont{Pollmann}},
  \bibinfo{author}{\bibfnamefont{A.~M.} \bibnamefont{Turner}},
  \bibinfo{author}{\bibfnamefont{E.}~\bibnamefont{Berg}}, \bibnamefont{and}
  \bibinfo{author}{\bibfnamefont{M.}~\bibnamefont{Oshikawa}},
  \bibinfo{journal}{Phys. Rev. B} \textbf{\bibinfo{volume}{81}},
  \bibinfo{pages}{064439} (\bibinfo{year}{2010}).

\bibitem[{\citenamefont{Chandran et~al.}(2013)\citenamefont{Chandran, Khemani,
  and Sondhi}}]{Chandran2013}
\bibinfo{author}{\bibfnamefont{A.}~\bibnamefont{Chandran}},
  \bibinfo{author}{\bibfnamefont{V.}~\bibnamefont{Khemani}}, \bibnamefont{and}
  \bibinfo{author}{\bibfnamefont{S.~L.} \bibnamefont{Sondhi}}
  (\bibinfo{year}{2013}), \bibinfo{note}{arXiv:1311.2946}.

\bibitem[{\citenamefont{You et~al.}(2014)\citenamefont{You, Bi, Rasmussen,
  Slagle, and Xu}}]{You2014}
\bibinfo{author}{\bibfnamefont{Y.-Z.} \bibnamefont{You}},
  \bibinfo{author}{\bibfnamefont{Z.}~\bibnamefont{Bi}},
  \bibinfo{author}{\bibfnamefont{A.}~\bibnamefont{Rasmussen}},
  \bibinfo{author}{\bibfnamefont{K.}~\bibnamefont{Slagle}}, \bibnamefont{and}
  \bibinfo{author}{\bibfnamefont{C.}~\bibnamefont{Xu}}, \bibinfo{journal}{Phys.
  Rev. Lett.} \textbf{\bibinfo{volume}{112}}, \bibinfo{pages}{247202}
  (\bibinfo{year}{2014}).

\bibitem[{\citenamefont{Sandvik and Kurkij\"arvi}(1991)}]{Sandvik1991}
\bibinfo{author}{\bibfnamefont{A.~W.} \bibnamefont{Sandvik}} \bibnamefont{and}
  \bibinfo{author}{\bibfnamefont{J.}~\bibnamefont{Kurkij\"arvi}},
  \bibinfo{journal}{Phys. Rev. B} \textbf{\bibinfo{volume}{43}},
  \bibinfo{pages}{5950} (\bibinfo{year}{1991}).

\bibitem[{\citenamefont{Sandvik}(1992)}]{Sandvik1992}
\bibinfo{author}{\bibfnamefont{A.~W.} \bibnamefont{Sandvik}},
  \bibinfo{journal}{Journal of Physics A: Mathematical and General}
  \textbf{\bibinfo{volume}{25}}, \bibinfo{pages}{3667} (\bibinfo{year}{1992}).

\bibitem[{\citenamefont{Sylju\aa{}sen and Sandvik}(2002)}]{Syljuasen2002}
\bibinfo{author}{\bibfnamefont{O.~F.} \bibnamefont{Sylju\aa{}sen}}
  \bibnamefont{and} \bibinfo{author}{\bibfnamefont{A.~W.}
  \bibnamefont{Sandvik}}, \bibinfo{journal}{Phys. Rev. E}
  \textbf{\bibinfo{volume}{66}}, \bibinfo{pages}{046701}
  (\bibinfo{year}{2002}).

\bibitem[{\citenamefont{Farhi et~al.}(2012)\citenamefont{Farhi, Gosset, Hen,
  Sandvik, Shor, Young, and Zamponi}}]{Farhi2012}
\bibinfo{author}{\bibfnamefont{E.}~\bibnamefont{Farhi}},
  \bibinfo{author}{\bibfnamefont{D.}~\bibnamefont{Gosset}},
  \bibinfo{author}{\bibfnamefont{I.}~\bibnamefont{Hen}},
  \bibinfo{author}{\bibfnamefont{A.~W.} \bibnamefont{Sandvik}},
  \bibinfo{author}{\bibfnamefont{P.}~\bibnamefont{Shor}},
  \bibinfo{author}{\bibfnamefont{A.~P.} \bibnamefont{Young}}, \bibnamefont{and}
  \bibinfo{author}{\bibfnamefont{F.}~\bibnamefont{Zamponi}},
  \bibinfo{journal}{Phys. Rev. A} \textbf{\bibinfo{volume}{86}},
  \bibinfo{pages}{052334} (\bibinfo{year}{2012}).

\bibitem[{\citenamefont{Haldane}(1983{\natexlab{a}})}]{Haldane1983a}
\bibinfo{author}{\bibfnamefont{F.}~\bibnamefont{Haldane}},
  \bibinfo{journal}{Phys. Lett. A} \textbf{\bibinfo{volume}{93}},
  \bibinfo{pages}{464 } (\bibinfo{year}{1983}{\natexlab{a}}).

\bibitem[{\citenamefont{Haldane}(1983{\natexlab{b}})}]{Haldane1983b}
\bibinfo{author}{\bibfnamefont{F.~D.~M.} \bibnamefont{Haldane}},
  \bibinfo{journal}{Phys. Rev. Lett.} \textbf{\bibinfo{volume}{50}},
  \bibinfo{pages}{1153} (\bibinfo{year}{1983}{\natexlab{b}}).

\bibitem[{\citenamefont{Affleck et~al.}(1987)\citenamefont{Affleck, Kennedy,
  Lieb, and Tasaki}}]{Affleck1987}
\bibinfo{author}{\bibfnamefont{I.}~\bibnamefont{Affleck}},
  \bibinfo{author}{\bibfnamefont{T.}~\bibnamefont{Kennedy}},
  \bibinfo{author}{\bibfnamefont{E.~H.} \bibnamefont{Lieb}}, \bibnamefont{and}
  \bibinfo{author}{\bibfnamefont{H.}~\bibnamefont{Tasaki}},
  \bibinfo{journal}{Phys. Rev. Lett.} \textbf{\bibinfo{volume}{59}},
  \bibinfo{pages}{799} (\bibinfo{year}{1987}).

\bibitem[{\citenamefont{den Nijs and Rommelse}(1989)}]{denNijs1989}
\bibinfo{author}{\bibfnamefont{M.}~\bibnamefont{den Nijs}} \bibnamefont{and}
  \bibinfo{author}{\bibfnamefont{K.}~\bibnamefont{Rommelse}},
  \bibinfo{journal}{Phys. Rev. B} \textbf{\bibinfo{volume}{40}},
  \bibinfo{pages}{4709} (\bibinfo{year}{1989}).

\bibitem[{\citenamefont{Pollmann et~al.}(2012)\citenamefont{Pollmann, Berg,
  Turner, and Oshikawa}}]{Pollmann2012}
\bibinfo{author}{\bibfnamefont{F.}~\bibnamefont{Pollmann}},
  \bibinfo{author}{\bibfnamefont{E.}~\bibnamefont{Berg}},
  \bibinfo{author}{\bibfnamefont{A.~M.} \bibnamefont{Turner}},
  \bibnamefont{and} \bibinfo{author}{\bibfnamefont{M.}~\bibnamefont{Oshikawa}},
  \bibinfo{journal}{Phys. Rev. B} \textbf{\bibinfo{volume}{85}},
  \bibinfo{pages}{075125} (\bibinfo{year}{2012}).

\bibitem[{\citenamefont{Todo et~al.}(2001)\citenamefont{Todo, Matsumoto,
  Yasuda, and Takayama}}]{Todo2001}
\bibinfo{author}{\bibfnamefont{S.}~\bibnamefont{Todo}},
  \bibinfo{author}{\bibfnamefont{M.}~\bibnamefont{Matsumoto}},
  \bibinfo{author}{\bibfnamefont{C.}~\bibnamefont{Yasuda}}, \bibnamefont{and}
  \bibinfo{author}{\bibfnamefont{H.}~\bibnamefont{Takayama}},
  \bibinfo{journal}{Phys. Rev. B} \textbf{\bibinfo{volume}{64}},
  \bibinfo{pages}{224412} (\bibinfo{year}{2001}).

\bibitem[{\citenamefont{Charrier et~al.}(2010)\citenamefont{Charrier, Capponi,
  Oshikawa, and Pujol}}]{Charrier2010}
\bibinfo{author}{\bibfnamefont{D.}~\bibnamefont{Charrier}},
  \bibinfo{author}{\bibfnamefont{S.}~\bibnamefont{Capponi}},
  \bibinfo{author}{\bibfnamefont{M.}~\bibnamefont{Oshikawa}}, \bibnamefont{and}
  \bibinfo{author}{\bibfnamefont{P.}~\bibnamefont{Pujol}},
  \bibinfo{journal}{Phys. Rev. B} \textbf{\bibinfo{volume}{82}},
  \bibinfo{pages}{075108} (\bibinfo{year}{2010}).

\bibitem[{\citenamefont{Degli Esposti~Boschi et~al.}(2003)\citenamefont{Degli
  Esposti~Boschi, Ercolessi, Ortolani, and Roncaglia}}]{DegliEspostiBoschi2003}
\bibinfo{author}{\bibfnamefont{C.}~\bibnamefont{Degli Esposti~Boschi}},
  \bibinfo{author}{\bibfnamefont{E.}~\bibnamefont{Ercolessi}},
  \bibinfo{author}{\bibfnamefont{F.}~\bibnamefont{Ortolani}}, \bibnamefont{and}
  \bibinfo{author}{\bibfnamefont{M.}~\bibnamefont{Roncaglia}},
  \bibinfo{journal}{Euro. Phys. J. B} \textbf{\bibinfo{volume}{35}},
  \bibinfo{pages}{465} (\bibinfo{year}{2003}).

\bibitem[{\citenamefont{Tzeng and Yang}(2008)}]{Tzeng2008a}
\bibinfo{author}{\bibfnamefont{Y.-C.} \bibnamefont{Tzeng}} \bibnamefont{and}
  \bibinfo{author}{\bibfnamefont{M.-F.} \bibnamefont{Yang}},
  \bibinfo{journal}{Phys. Rev. A} \textbf{\bibinfo{volume}{77}},
  \bibinfo{pages}{012311} (\bibinfo{year}{2008}).

\bibitem[{\citenamefont{Tzeng et~al.}(2008)\citenamefont{Tzeng, Hung, Chen, and
  Yang}}]{Tzeng2008b}
\bibinfo{author}{\bibfnamefont{Y.-C.} \bibnamefont{Tzeng}},
  \bibinfo{author}{\bibfnamefont{H.-H.} \bibnamefont{Hung}},
  \bibinfo{author}{\bibfnamefont{Y.-C.} \bibnamefont{Chen}}, \bibnamefont{and}
  \bibinfo{author}{\bibfnamefont{M.-F.} \bibnamefont{Yang}},
  \bibinfo{journal}{Phys. Rev. A} \textbf{\bibinfo{volume}{77}},
  \bibinfo{pages}{062321} (\bibinfo{year}{2008}).

\bibitem[{\citenamefont{Albuquerque et~al.}(2009)\citenamefont{Albuquerque,
  Hamer, and Oitmaa}}]{Albuquerque2009}
\bibinfo{author}{\bibfnamefont{A.~F.} \bibnamefont{Albuquerque}},
  \bibinfo{author}{\bibfnamefont{C.~J.} \bibnamefont{Hamer}}, \bibnamefont{and}
  \bibinfo{author}{\bibfnamefont{J.}~\bibnamefont{Oitmaa}},
  \bibinfo{journal}{Phys. Rev. B} \textbf{\bibinfo{volume}{79}},
  \bibinfo{pages}{054412} (\bibinfo{year}{2009}).

\bibitem[{\citenamefont{Hu et~al.}(2011)\citenamefont{Hu, Normand, Wang, and
  Yu}}]{Hu2011}
\bibinfo{author}{\bibfnamefont{S.}~\bibnamefont{Hu}},
  \bibinfo{author}{\bibfnamefont{B.}~\bibnamefont{Normand}},
  \bibinfo{author}{\bibfnamefont{X.}~\bibnamefont{Wang}}, \bibnamefont{and}
  \bibinfo{author}{\bibfnamefont{L.}~\bibnamefont{Yu}}, \bibinfo{journal}{Phys.
  Rev. B} \textbf{\bibinfo{volume}{84}}, \bibinfo{pages}{220402}
  (\bibinfo{year}{2011}).

\bibitem[{\citenamefont{Pollmann and Turner}(2012)}]{Pollmann2012b}
\bibinfo{author}{\bibfnamefont{F.}~\bibnamefont{Pollmann}} \bibnamefont{and}
  \bibinfo{author}{\bibfnamefont{A.~M.} \bibnamefont{Turner}},
  \bibinfo{journal}{Phys. Rev. B} \textbf{\bibinfo{volume}{86}},
  \bibinfo{pages}{125441} (\bibinfo{year}{2012}).

\bibitem[{\citenamefont{Wierschem and Sengupta}(2014)}]{Wierschem2014}
\bibinfo{author}{\bibfnamefont{K.}~\bibnamefont{Wierschem}} \bibnamefont{and}
  \bibinfo{author}{\bibfnamefont{P.}~\bibnamefont{Sengupta}},
  \bibinfo{journal}{Phys. Rev. Lett.} \textbf{\bibinfo{volume}{112}},
  \bibinfo{pages}{247203} (\bibinfo{year}{2014}).

\end{thebibliography}

\end{document}